\documentclass[prb,twocolumn]{revtex4}
\usepackage{bm}
\usepackage{graphicx}
\usepackage{amssymb}
\usepackage{amsmath}
\usepackage{eufrak}
\usepackage{color}
\usepackage[utf8]{inputenc} 
\usepackage{hyperref}
\usepackage{pifont}
\usepackage{ulem}
\usepackage{array}
\usepackage{verbatim} 
\usepackage{placeins}

\newcommand{\nix}[1]{}

\begin{document}
	
	\title{Opto-Electronic Characterization of Three Dimensional Topological Insulators}
	
	\author{
		H.~Plank$^1$, S.\,N.~Danilov$^1$, 
		V.\,V.~Bel’kov$^1$\footnote{permanent address: Ioffe Physical-Technical Institute, St. Petersburg, Russia}, 
		V.\,A.~Shalygin$^1$\footnote{permanent address: Saint-Petersburg State Polytechnic University, St. Petersburg, Russia}, 
		J.~Kampmeier$^2$, M.~Lanius$^2$, G.~Mussler$^2$, D.~Gr\"{u}tzmacher$^2$, 
		and S.\,D.~Ganichev$^1$	
		}
	\affiliation{$^1$Terahertz Center, University of Regensburg, Regensburg, Germany}
	\affiliation{$^2$Peter Gr\"unberg Institute (PGI) \& J\"ulich Aachen Research Alliance (JARA-FIT) J\"ulich, Germany}
\begin{abstract}	
We demonstrate that the terahertz/infrared radiation induced photogalvanic effect, which is sensitive to the surface symmetry and scattering details, can be applied to study the high frequency conductivity of the surface states in (Bi$_{1-x}$Sb$_{x}$)$_{2}$Te$_{3}$ based three dimensional (3D) topological insulators (TI). 
In particular, measuring the polarization dependence of the photogalvanic current and scanning with a micrometre  sized beam spot across the sample, provides access to (i)  
topographical inhomogeneity's  in the electronic properties of the surface states and (ii) the local domain orientation.  
An important advantage of the proposed method is that it can be applied to study TIs at room temperature and even in materials with a high electron density of bulk carriers.
\end{abstract}
	
\maketitle{} 

Electronic, optical and opto-electronic properties  of topological insulators (TI)  have attracted continuously growing attention yielding challenging fundamental concepts and being of potential interest for novel applications in the fields of spintronics and opto-electronics~\cite{HasanKane2010, Moore2010, QiZhang2011, Bardarson2013, Bernevig2013}. 
Hence, the fabrication of high quality topological insulators and their characterization yielding feedback to technologists is of particular importance. 
Until now a large variety of materials was proposed and confirmed to host topological protected surface states in three-dimensional (3D) TI and edge channels in two-dimensional (2D) TI. 
Particular examples are (Bi$_{1-x}$Sb$_{x}$)$_{2}$Te$_{3}$ based 3D TI.
Their fabrication in view of good insulating properties of the  bulk at room temperature, homogeneity of a large area growth materials is  still a challenging task.
The former problem is caused by the high density of residual impurities serving parallel channels to the surface transport~\cite{bulk_states1, bulk_states2, bulk_states3}. 
A promising way to overcome this problem serves the recent progress in growth of 3D TI applying molecular-beam-epitaxy (MBE) technique, see e.g.,~[\onlinecite{MBE_1, MBE_2, MBE_3, MBE_4}]. 
Owing to the progress in material growth,  low temperature electron transport and magneto-transport studies becomes possible providing  information on average electronic properties of Dirac fermions and carrier scattering  mechanisms in 3D TIs~\cite{transport1, transport2, transport3, transport4, transport5, transport6, Barreto2014, Ren2010, Qu2010} as well as to observe  the quantum anomalous Hall effect reported for Cr or V doped (Bi$_{1-x}$Sb$_{x}$)$_{2}$Te$_{3}$ based 3D TI~\cite{Trans_doped1, Trans_doped2, Trans_doped3, Trans_doped4}. 
An important issue for improvement of the material properties is their characterization allowing insights in the material properties and providing a feedback for technologists. For that a palette of methods has been developed and widely used.
An insight into the band structure of the surface states of 3D TIs, especially proof for the single Dirac cone, 
is obtained by varios modifications of the angle resolved photoemission spectroscopy (ARPES)~\cite{ARPES_2, ARPES_3, ARPES_5, ARPES_1, ARPES_4, timeARPES, ARPES_6}, including  spin-resolved and time-resolved ARPES,
as well as by time-resolved two-photon photoelectron (2PPE) spectroscopy~\cite{2ppe1, 2ppe3, 2ppe2}, 
with which an enlightening pictures of fast dynamics in carrier relaxation can be obtained. 
\\
Further methods, providing important information concerning different growth parameters, nucleation of trigonal islands, and domain alignment of the TI film with respect to the substrate, include  scanning  (STEM) and high angle annular dark field (HAADF) transmission electron microscopy~\cite{STM1, STM2, STM3, Samarath}; scanning electron microscope (SEM)~\cite{SEM};  atomic force microscopy (AFM)~\cite{AFM}; $X$-ray diffraction (XRD)~\cite{XRD}; and second harmonic generation~\cite{Hamh} as well as infrared or optical spectroscopy~\cite{Basov,Loosdrecht}. 
However, both spectroscopy and transport investigations do not allow one to analyze the local conductivity in particular at room temperature and materials homogeneity on a large scale.

Here we demonstrate that the study of terahertz radiation induced electron transport  in (Bi$_{1-x}$Sb$_{x}$)$_{2}$Te$_{3}$ based 3D TI  provides experimental access to a spatial resolved characterization of  transport properties of Dirac fermions even at room temperature and allows one to map the domain orientation at different locations. 
The suggested method combines a 2D scanning technique with the detection of the linear photogalvanic effect (LPGE)\cite{Ganichev2003,book}
recently observed in various TI materials~\cite{Olbrich2014,Kastl2015_1,Braun2015,Plank2016}. 
Due to symmetry filtration discussed below, the LPGE is only excited in topological surface states even in materials with a high bulk carrier density for which  \textit{dc} electron transport characterization fails. 
Owing to the fact that the sign and the strength of the effect are determined by the domain orientation and the processes of momentum relaxation,  the detection of the LPGE current, locally excited at different points on TI film, makes  possible to map the domains arrangement, to judge on the presence of twin domains~\cite{Samarath} in the samples, to study high frequency conductivity and to estimate electron scattering times of the surface carriers.

\begin{figure}
	\includegraphics[width=\linewidth]{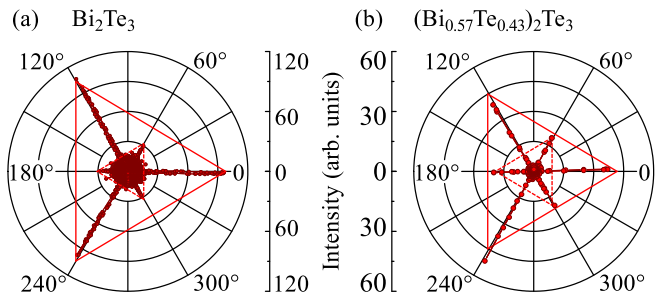}
	\caption{
			X-ray diffraction data of studied Bi$_2$Te$_3$ (a) and (Bi$_{0.57}$Sb$_{0.43}$)$_2$Te$_3$ (b) samples.	
		}
	\label{figure1xrays}
\end{figure}

\begin{figure}
	\includegraphics[width=\linewidth]{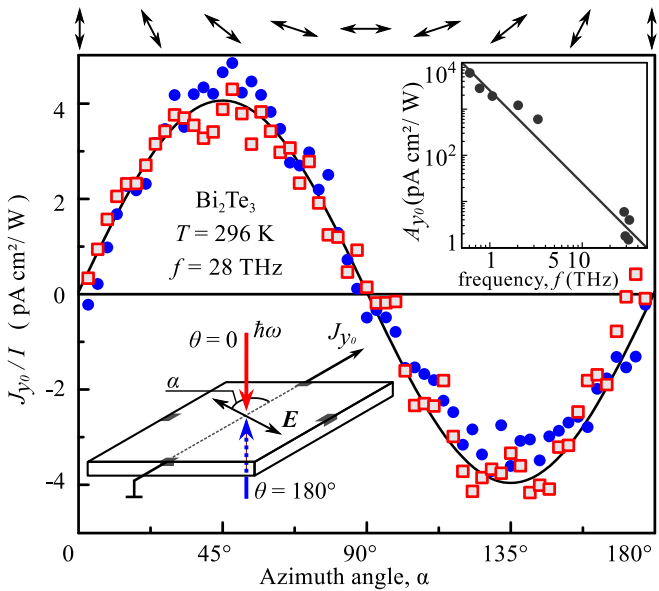}
	\caption{
		Dependence of the photocurrent normalized by the radiation intensity $J_{y_0} / I$ on the azimuth angle $\alpha$ in a Bi$_2$Te$_3$ sample, illuminated at front and back at 28 THz.
		Solid line shows fit after Eq.~(\ref{PGE}). 
		Arrows on the top indicate electric field orientation for several angles $\alpha$.
		Insets show experimental setup and frequency dependence of the photocurrent magnitude $A_{y_0}$ measured in the frequency range $f=0.6$ - 32.6~THz. 
		The line shows fit after Eqs.~(\ref{PGE}) and (\ref{sigma}) demonstrating that in the studied frequency range the photocurrent amplitude varies as $A \propto f^{-2}$. 
		}
	\label{figure2pge}
\end{figure}

The experiments were carried out on a Bi$_2$Te$_3$ and a (Bi$_{0.57}$Sb$_{0.43}$)$_{2}$Te$_{3}$ samples grown on a silicon (111) substrates by means of van der Waals epitaxy, where weak bonds between substrate and the TI epilayers reduce the strength and therefore the large lattice mismatch does not hinder the growth of single crystal TI films with a high structural quality~\cite{Borisova2012, Plucinski2013, Borisova2013, Kampmeier2015, ternaries}. 
All samples are $n$-type with bulk carrier densities in the order of  $n = 5 \times 10^{19}$~cm$^{-3}$ and $n = 3 \times 10^{17}$~cm$^{-3}$ in Bi$_2$Te$_3$  and (Bi$_{0.57}$Sb$_{0.43}$)$_{2}$Te$_{3}$, respectively. 
The existence of surface states with linear dispersion and the energetic position of the Dirac point, $E_{\rm DP}$, with respect to the Fermi energy has been proved in all studied samples by ARPES ~\cite{ARPES_2,ARPES_3}. In both samples the energy of the Dirac point $E_{\rm DP}$ is in the order of hundreds of millielectronvolts.
Additionally, XRD measurements were performed in order to verify the single~-~crystallinity of the thin films and to determine the domain orientation. 
The XRD data, see Fig.~\ref{figure1xrays}, demonstrate the formation of two types of trigonal domains, being mirror-symmetric to each other and show, however, that the majority of the domains have the same orientation~\cite{Olbrich2014,Kampmeier2015} with wedges heights (altitudes) and  bases being aligned along the crystallographic axes of Si substrate ${x_0} \parallel [11{\bar 2}]$ and $y_0 \parallel [{\bar 1}10]$, respectively. 
Figures~\ref{figure1xrays} (a) and (b) show that the ratio between the dominant domain and the suppressed one is in the order of 3.5 Bi$_2$Te$_3$ and 2.1 (Bi$_{0.57}$Sb$_{0.43}$)$_{2}$Te$_{3}$, respectively. 
Using the XRD results we prepared rectangle shaped samples with edges cut along $x_0$ and $y_0$.
To enable electrical measurements two pairs of ohmic contacts have been prepared in the middles of the $4\times7$\,mm$^2$ sample's edges, so that the terahertz radiation induced \textit{dc} current $\bm j$ is probed along the crystallographic axes.
Further information on the surface domains have been obtained by high-resolution STEM and AFM, demonstrating that trigonal islands with quintuple layers (QLs) step heights of about 1\,nm nucleate on the surface and follow the orientation of the domains~\cite{Olbrich2014,Borisova2013}.
Images obtained by scanning electron microscopy (SEM) show the presence of islands with a peak to valley roughness of ~ 25$\%$ of the nominal thickness.
%
To excite photocurrents
we applied radiation of pulsed line-tunable mid-infrared  $Q$-switched and TEA CO$_2$ 
lasers~\cite{removalSiGe,Resonantinversion2003,SGEopt2003,Chongyun}, 
as well as an optically pumped molecular terahertz (THz) 
laser~\cite{Lechner2009,Drexler2013,ratchet2009,helix2012}.
With the CO$_2$ lasers we examined photocurrents in the frequency range between  $f=28$ and 32.6~THz (photon energies $\hbar \omega$ from 114  to 135~meV) and with the THz laser in the frequency range between 0.6 and 3.9~THz ($\hbar \omega$ from 2.5 to 16~meV).
The lasers provided 
pulse durations in the order of 100~ns and peak power ranging from 1 to 10 kW. 
The radiation power was controlled by mercury-cadmium-telluride and photon drag detectors~\cite{Ganichev84p20}. 
The polarization plane orientation of linearly polarized light is described by an azimuth angle $\alpha$. 
To measure the photocurrent as a function of angle $\alpha$ we rotated
a $\lambda/2$ plate or a linear polarizer placed behind a Fresnel $\lambda/4$ rhomb
~\cite{BelkovSSTlateral,ratchet2011}.
The radiation was focused on the samples with a spot size diameters for the THz laser of 2~mm and for the CO$_2$ lasers in the range of 0.5~mm down to several tens of $\mu$m. 
The radiation profile was measured applying a pyroelectrical camera~\cite{Ganichev1999,Ziemann2000},
confirming an almost Gaussian form of the beam. 
To carry out spatially resolved measurements the beam was scanned across the sample. 
The highest spatial resolution used here to study the distribution of the photocurrents magnitude across the sample
was obtained applying mid-infrared lasers, which were focused into a spot of about 30~$\mu$m.
The photocurrent excited in unbiased samples at room temperature was recorded with a 1~GHz storage oscilloscope. 
The experimental setup is sketched in the inset in Fig.~\ref{figure2pge}.
The samples were illuminated at normal incidence with front and back illumination, with corresponding angle of incidence $\theta=0$ and $180^\circ$, respectively.

Photocurrents are observed in the whole frequency range from 0.6 to 32.6 THz. 
The signal follows the temporal structure of the laser pulse intensity and scales linearly with the radiation intensity. 
It exhibits a characteristic dependence upon the rotation of the $ac$ electric field orientation as $J_{{x_0}}= - A(f) \cos{2 \alpha}$ and 
$J_{y_0}(\alpha) = A(f) \sin{2 \alpha}$. This behaviour is shown exemplary in Fig.~\ref{figure2pge} 
for the photocurrent excited in the Bi$_2$Te$_3$ sample measured in $y_0$-direction and excited by radiation with $f=28$~THz.  
Cooling the sample from 296 to 4.2~K increases the magnitude of the photocurrent, but the polarization dependence is retained. 
Varying the $ac$-electric field frequency we obtained that the parameter $A(f)$, which determines the photocurrent magnitude, strongly increases with the frequency decrease.
The inset in Fig.~\ref{figure2pge} shows that $A(f)$ closely follows the law $A \propto 1/f^2$ in the whole range of studied frequencies. 
The  photocurrents demonstrating such behaviour can be caused either by the linear photogalvanic or 
orby the photon drag effect, as it was observed early in Bi$_2$Te$_3$ and (Bi$_{0.57}$Sb$_{0.43}$)$_{2}$Te$_{3}$ in the low frequency range~\cite{Olbrich2014,Plank2016}.
An experiment applying back and front illumination allows us to judge on the mechanism dominating the current formation in the studied samples. 
Indeed, while the LPGE being determined by the \textit{in-plane} orientation of the radiation electric field~\cite{Olbrich2014} only remains unchanged for both geometries, the transvers  photon drag effect, being additionally proportional to the photon momentum $q_z$, must invert the sign of the photocurrent since ${q_z} \rightarrow - {q_z}$  is changing from front to back illumination. 
For all studied frequencies and samples we observed that the data for both geometries almost coincide, see Fig.~\ref{figure2pge} for $f=28$~THz in the Bi$_2$Te$_3$ sample, confirming that the observed photocurrent is caused by the linear photogalvanic effect. 
A specific feature of the photogalvanic effect, whose prerequisite is the lack of an inversion centre, is that in our centrosymmetric 3D TI materials it is excited in the non-centrosymmetric surface states only. 
For all frequencies used, the photon energies $\hbar \omega$ are substantially smaller than the Fermi energy $E_{\rm F}$, which is introduced as the energy difference between $E_{\rm DP}$ and the highest occupied state, 
as confirmed by ARPES measurements~\cite{Plank2016}. 
Consequently, the photocurrent is caused by the Drude-like free carrier absorption. 
The trigonal symmetry of 2D surface carriers (the point group of the surface is C$_{\rm 3v}$) makes the elastic scattering asymmetric. 
For the surface states the scatterers can be considered as randomly distributed but identically oriented wedges lying in the QL-planes. 
The preferential orientation of wedges is supported by the $X$-rays data shown above, see Fig.~\ref{figure1xrays} and  Refs.~[\onlinecite{Olbrich2014,NatPhys09}]. 
for details.
%
%
\begin{figure}
	\includegraphics[width=\linewidth]{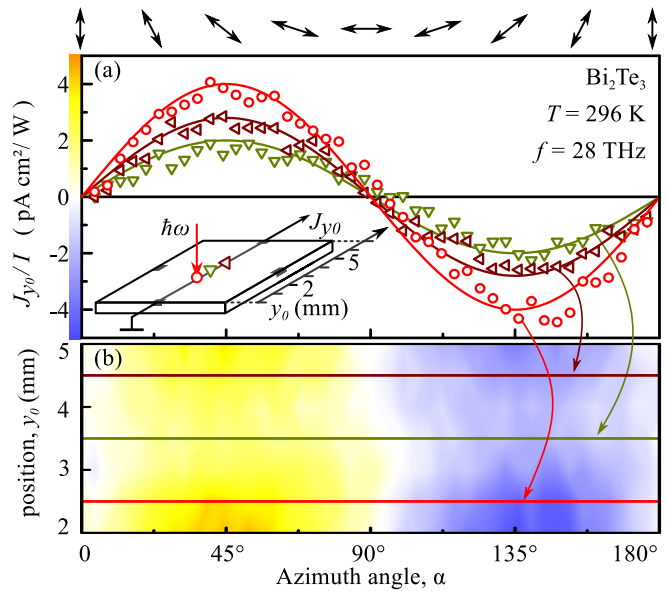}
	\caption{
		(a) Dependence of the photocurrent normalized on the radiation intensity $J_{y_0} / I$ on the azimuth angle $\alpha$ measured in a Bi$_2$Te$_3$ sample for three different laser spot positions, see inset.
		Solid lines show fit after Eq.~(\ref{PGE}) and arrows on the top indicate electric field orientation for 
		several angles $\alpha$. 
		(b) Colour-coded photocurrent strength as a function of the angle $\alpha$ obtained by scanning across the sample parallel to $y_0$-axis with the step of $500~\mu$m. 
	}
	\label{figure3scan1}
\end{figure}
%
Application of linearly polarized THz radiation causes an \textit{alignment} of carrier momenta: the total flow of electrons driven back and forth by the \textit{ac} electric field $\bm{E}(t)$ increase. 
The corresponding \textit{stationary} correction to the electron distribution function scales as a square of the $ac$ electric field magnitude. 
Due to asymmetric scattering by wedges, the excess of the flux of Dirac fermions moving back and forth along the external $ac$-electric field results in a $\textit{dc}$ photogalvanic current, for details see Refs.~\cite{Olbrich2014, Plank2016}. 
The direction and magnitude of the photocurrent depends
 on the orientation of the wedges with
respect to the $x_0-$ and the $y_0-$~directions, the in-plane orientation of the \textit{ac} electric field vector,   as well as on details of carrier scattering. For arbitrary orientation of domains in respect to the $x_0/y_0$ directions in which the photocurrent is measured  
and elastic scattering by Coulomb impurities~\cite{footnoteT1} the photogalvanic current is given by~\cite{Olbrich2014} (see Suppl. Mater.) 
\begin{align}
	j_{x_0} = -\cos(2\alpha - 3\Phi_0)  e v_0 (\sigma(\omega)/ E_F) \tau_{tr} \Xi   |E_0|^2\nonumber, \\
	j_{y_0} = +\sin(2\alpha - 3\Phi_0)  e v_0 (\sigma(\omega)/ E_F) \tau_{tr} \Xi   |E_0|^2, 
	\label{PGE}
\end{align}
where $e$ is the electron charge, $v_0$ is the Fermi velocity, $\Phi$ is angle between $x_0$ and one of the mirror reflection plane of wedges, $\sigma(\omega)$ is the high-frequency conductivity given by the Drude expression 
\begin{equation}
	\sigma(\omega) = {e^2 E_F \tau_\text{tr} \over 4\pi\hbar^2 [1+(\omega\tau_{\rm tr})^2]} \, ,
	\label{sigma}
\end{equation}
where $\tau_{tr} = 3 \tau_{2} \propto E_{\rm F}$ is the transport scattering time and $|E_0|^2$ is the squared magnitude of the \textit{ac}  electric field.  
The asymmetry of the scattering probability is given by $\Xi = \tau_{tr}\sum_{p'}<2\cos\varphi_{p}\cos2\varphi_{p'}W^{(a)}_{p',p}>_{\varphi_p}$ 
where $W^{(a)}_{p',p}$ is the asymmetric probability for carrier to have the momentum $p$ ($p'$) and the polar angle $\varphi_{p}$ ($\varphi_{p'}$)  before (after) a scattering event. 
For $\Phi = 0$ the above expression describes the photocurrent excited in the directions parallel and perpendicular to one of the mirror reflection planes of the C$_{\rm 3v}$ point group. 
%
\begin{figure}
	\includegraphics[width=\linewidth]{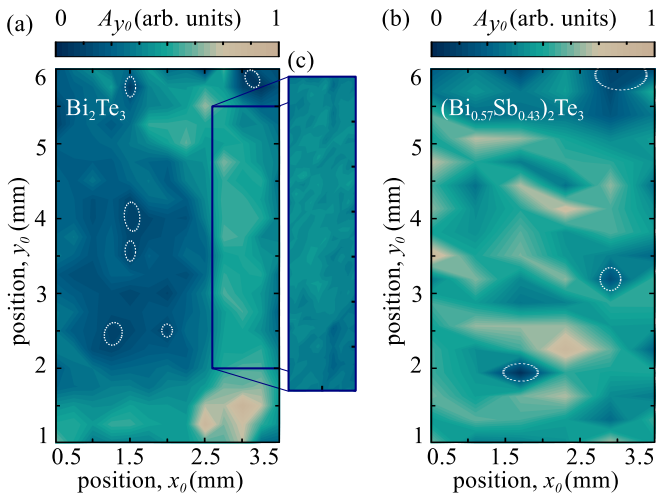}
	\caption{
		Colour-coded photocurrent amplitude $A_{y_0}$ as a function of coordinate, obtained for $f = 28$~THz and beam spot of about 30~$\mu$m by a two-dimensional scan in two samples (a) Bi$_2$Te$_3$ and 
		(b) (Bi$_{0.57}$Sb$_{0.43}$)$_{2}$Te$_{3}$. 
		The scans are obtained in a rough scan mode with steps of $250~\mu$m. 
		Areas with vanishingly small photocurrent signals are indicated by white circles. 
		Panel (c) shows the fine scan image obtained with steps of $100~\mu$m. 
		All data are obtained at room temperature. 
	}
	\label{figure4scan2}
\end{figure}
Equations~(\ref{PGE}) and (\ref{sigma}) show that the frequency dependence of the photocurrent is 
determined by that of the high-frequency conductivity, $\sigma(\omega)$. 
The fact that the amplitude of the photocurrent scales as $\omega^{-2}$, see inset in Fig.~\ref{figure2pge},  reveals that $\omega \tau_\text{tr}$ exceeds unity for all applied frequencies.  
Therefore, the room temperature electron scattering times of the surface carriers $\tau_{\rm tr}$ in our samples is at least larger than 0.2~ps. 
Taking into account the value of the Fermi energy measured by ARPES\cite{Plank2016}, we estimate the lowest limit of the surface electrons mobility~\cite{footnoteT3} $\mu \approx 1.5 \times 10^{3}$\,cm$^2$/Vs and $0.97 \times 10^{3}$\,cm$^2$/Vs at $T = 296$~K for Bi$_2$Te$_3$ and (Bi$_{0.57}$Sb$_{0.43}$)$_{2}$Te$_{3}$ samples, respectively.
Furthermore, in the limit of $\omega \tau_\text{tr} \gg 1$  both, scattering time and Fermi energy cancel in the formulas for $j_{x_0}$ and $j_{y_0}$, see Eq.~\ref{PGE}) and Eq.~(\ref{sigma}), and we derive the amplitude of the photocurrent normalized by the radiation power as $A = e^3 v_0 \Xi  / 4\pi\hbar^2 \omega^2$.
Using $A$ from the inset in Fig.~\ref{figure2pge} and the Fermi velocity $v_0 = 5.1 \times 10^{5}$~m/s for Bi$_2$Te$_3$ and $3.8 \times 10^{5}$~m/s for (Bi$_{0.57}$Sb$_{0.43}$)$_{2}$Te$_{3}$ we obtain the asymmetric probability of a carrier scattering. 
Note that Fermi velocities in both samples are determined from the ARPES data presented in~\cite{Plank2016}
 and give the following values of the asymmetry of the scattering probability: $\Xi = 4.3 \times 10^{-3}$ for Bi$_2$Te$_3$ and $\Xi = 1.4 \times 10^{-3}$ for (Bi$_{0.57}$Sb$_{0.43}$)$_{2}$Te$_{3}$. 
Moreover, using Eq.~(\ref{PGE}) one can analyse the domain orientation and conclude on the possible presence of twisted surface domains. 
Figure~\ref{figure3scan1}(a) shows the polarization dependence of the photocurrent excited in the Bi$_2$Te$_3$ sample for three different beam spot positions along the $y_0$-direction. 
Fig.~\ref{figure3scan1}(b) presents the results of a scan along the same line. 
The data obtained with a half millimetre sized laser spot yield information on the average domain orientation. 
At all positions we obtained the same phase angle $\Phi = 0$, which is in well agreement with the XRD data, see Fig.~\ref{figure1xrays} (a). The magnitude of the photocurrent along this line, as long as contacts or the sample edges are not illuminated~\cite{footnoteT2}, varies by about 60\%. 
In the second sample (data not shown) we also detected a phase angle $\Phi = 0$, being in agreement with the XRD data shown in Fig.~\ref{figure1xrays} (b), and similar variation of the photocurrent magnitude from one point to another.
As in these measurements the photocurrent is averaged over the large laser spot size, it is not sensitive to small deviations, which could be caused by defects in the material or ad-atoms on the surface. They also do not reflect possible local misalignments/disorientations of the domains and the presence of the twin domains. Information on the topography of the domain orientation and electron transport properties can, however, be obtained by focusing the radiation into a smaller spot size.
Figure~\ref{figure4scan2} presents corresponding two dimensional scans measured in Bi$_2$Te$_3$ (a) and   (Bi$_{1-x}$Sb$_{x}$)$_{2}$Te$_{3}$ (b) samples with a beam spot of about 30~$\mu$m.
Although the amplitude of the photocurrent deviates much stronger from point to point than that measured with a larger beam spot, 
the LPGE signal is still observed at all beam positions.
In both samples and for all laser spot positions neither arbitrary values of the phase angle $\Phi$ nor switching to opposite domain orientation (for $\Phi = 180^\circ$ photocurrent should change its sign) have been detected~-~the phase angle $\Phi$ remains zero. 
However, for a few laser spot positions we registered vanishingly small photocurrent signals, these points are indicated in Fig.~\ref{figure4scan2} by white circles. 
These almost zero photocurrents, indicate the presence of twisted domains for which the currents generated by mirror-symmetric domains cancel each other ($j^{\Phi=0} = - j^{\Phi=180^\circ}$). 
Further analysis of the photocurrent magnitude distributions obtained on Bi$_2$Te$_3$ and (Bi$_{1-x}$Sb$_{x}$)$_{2}$Te$_{3}$ samples reveals that the latter one shows more inhomogeneous transport properties than the former one.
This observation agrees well with the fact known for technologists, that pure Bi$_2$Te$_3$ is usually growing smooth compared to Sb$_2$Te$_3$ ones. By adding antimony to the binary Bi$_2$Te$_3$ and growing (Bi$_{1-x}$Sb$_{x}$)$_{2}$Te$_{3}$, more rough the sample surface gets.
To examine the most homogeneous part with the highest signals in the Bi$_2$Te$_3$ sample indicated by rectangle in Fig.~\ref{figure4scan2}(a) we used a fine step mode with the scan step of $100~\mu$m. 
Results, shown on the right hand side of panel (a), indicate that the room temperature surface transport can be almost homogeneous even within a very large area of 3.5$\times$1.0~mm$^2$.  

To summarize, our study shows that the linear photogalvanic effect provides an efficient tool to study details of the sample homogeneity and the surface transport in 3D TIs. 
A particular advantage of the method is that it can be applied in a wide temperature range, including room temperature being of importance for technology
as well as in samples with a high bulk carrier density caused by residual impurities, i.e. under the conditions which 
almost exclude conventional transport methods.  
Moreover, being demonstrated here for Bi$_2$Te$_3$-based materials it can also be applied to other 3D TI.
\acknowledgments
We thank Leonid~E. Golub and Johannes Ziegler for fruitful discussions. 
The support from the DFG priority program SPP1666,
the Elite Network of Bavaria (K-NW-2013-247), and 
the Programm of the VW Stiftung
is gratefully acknowledged.


\begin{thebibliography}{99}
	
	\bibitem{HasanKane2010} M. Z. Hasan and C. L. Kane, 
	Rev. Mod. Phys. \textbf{82}, 3045 (2010).
	
	\bibitem{Moore2010} J. E. Moore, 
	Nature \textbf{464}, 194 (2010).
	
	\bibitem{QiZhang2011}  X. L. Qi and S. C. Zhang,
	Rev. Mod. Phys. \textbf{83}, 1057 (2011).
	
	\bibitem{Bardarson2013} J. H. Bardarson and J. E. Moore, 
	Rep. Prog. Phys. \textbf{76}, 056501 (2013). 
	
	\bibitem{Bernevig2013} B. A. Bernevig, 
	\textit{Topological Insulators and Superconductors}
	(University Press Group Ltd, 2013).
	
	
	\bibitem{bulk_states1} J. G. Checkelsky, Y. S. Hor, M. H. Liu, D. X. Qu, R. J. Cava, and N. P. Ong,
	Phys. Rev. Lett. \textbf{103}, 246601 (2009).
	
	\bibitem{bulk_states2} A. A. Taskin and Y. Ando,
	Phys. Rev. B \textbf{80}, 085303 (2009).
	
	\bibitem{bulk_states3}  J. G. Analytis, J. H. Chu, Y. Chen, F. Corredor, R. D. McDonald, Z. X. Shen, and I. R. Fisher,
	Phys. Rev. B \textbf{81}, 205407 (2010). 
	
	\bibitem{MBE_1} L. He, X. Kou, K. L. Wang, 
	Phys. Status Solidi RRL \textbf{7}, 50 
	(2013).
	
	\bibitem{MBE_2} G. Wang, X.-G. Zhu, Y.-Y. Sun, Y.-Y. Li, T. Zhang, J. Wen, X. Chen, K. He, L.-L. Wang, X.-C. Ma, J.-F. Jia, S. B. Zhang, and Q.-K.  Xue, 
	Adv. Mater. \textbf{23}, 2929 (2011).
	
	\bibitem{MBE_3} J. Krumrain, G. Mussler, S. Borisova, T. Stoica, L. Plucinski, C. M. Schneider, and D. Grützmacher, 
	J. Crystal Growth \textbf{324}, 115  (2011).
	
	\bibitem{MBE_4} S. E. Harrison, S. Li, Y. Huo, B. Zhou, Y. L. Chen, and J. S. Harris
	Appl. Phys. Lett. \textbf{102}, 171906 (2013).
	
	\bibitem{transport1}  T. C. Hsiung, C. Y. Mou, T.K. Lee, and Y. Y. Chen
	Nanoscale \textbf{7}, 518 (2015).
	
	\bibitem{transport2}  J. Yong, Y. P. Jiang, D. Usanmaz, S.  Curtarolo, X. H. Zhang, L. Z. Li, X. Q.  Pan, J.  Shin, I. Takeuchi, and R. L. Greene
	Appl. Phys. Lett. \textbf{105}, 222403 (2014).
	
	\bibitem{transport3} T. C. Hsiung, D. Y. Chen, L.  Zhao, Y. H.  Lin, C. Y.  Mou, T. K.  Lee, M. K.  Wu, and Y. Y. Chen
	Appl. Phys. Lett \textbf{103}, 163111 (2013).
	
	\bibitem{transport4} B. Xia, P. Ren, A. Sulaev, P. Liu, S. Q.  Shen, and L. Wang
	Phys. Rev. B \textbf{87}, 085442 (2013).
	
	\bibitem{transport5} Z. Ren, A. A.  Taskin, S.  Sasaki, K. Segawa, and Y. Ando
	Phys. Rev. B. \textbf{85}, 155301 (2012).
	
	\bibitem{transport6} L. He, F. X.  Xiu, X. X.  Yu, M. Teague, M W. J. Jiang, Y. B.  Fan, X. F.  Kou, M. R. Lang, Y.  Wang, G.  Huang, N. C.  Yeh, and K. L. Wang
	Nano Lett. \textbf{12}, 1486 (2012).
	
	\bibitem{Barreto2014} L. Barreto, L. K\"{u}hnemund, F. Edler, Ch. Tegenkamp, J. Mi, M. Bremholm, B. B. Iversen, Ch. Frydendahl, M. Bianchi, and P. Hofmann
	Nano Lett. \textbf{14}, 3755
	(2014).
	
	\bibitem{Ren2010} Z. Ren, A. A. Taskin, S. Sasaki, K. Segawa, and Y. Ando
	Phys. Rev. B \textbf{82}, 241306(R) (2010).
	
	\bibitem{Qu2010} D.-X. Qu, Y. S. Hor, J. Xiong, R. J. Cava, N. P. Ong
	Science \textbf{329}, 821 (2010).
	
	\bibitem{Trans_doped1} C.-Z. Chang, J. Zhang, X. Feng, J. Shen, Z. Zhang, M. Guo, K. Li, Y. Ou, P. Wei, L.-L. Wang, Z.-Q. Ji, Y. Feng, S. Ji, X. Chen, J. Jia, X. Dai, Z. Fang, S.-C. Zhang, K. He, Y. Wang, L. Lu, X.-C. Ma, and Q.-K. Xue
	Science  \textbf{340}, 6129 (2013).
	
	\bibitem{Trans_doped2} C.-Z. Chang, W. Zhao, D. Y. Kim, H. Zhang, B. A. Assaf, D. Heiman, S.-C. Zhang, C. Liu, M. H. W. Chan, and  J. S. Moodera
	Nature Materials \textbf{14}, 473
	(2015).
	
	\bibitem{Trans_doped3} A. J. Bestwick, E. J. Fox, X. Kou, L. Pan, K. L. Wang, and D. Goldhaber-Gordon
	Phys. Rev. Lett. \textbf{114}, 187201 (2015).
	
	\bibitem{Trans_doped4} A. Kandala,  A. Richardella,  S. Kempinger,  C.-X. Liu, and N. Samarth
	Nat. Comm. \textbf{6}, 7434 (2015).
	
	\bibitem{ARPES_2} Y. Xia, D. Qian, D. Hsieh, L. Wray, A. Pal, H. Lin, A. Bansil, D. Grauer, Y. S. Hor, R. J. Cava, and M. Z. Hasan,
	Nature Physics \textbf{5}, 398 (2009).
	
	\bibitem{ARPES_3}  D. Hsieh, Y. Xia, D. Qian, L. Wray, J. H. Dil, F. Meier, J. Osterwalder, L. Patthey, J. G. Checkelsky, N. P. Ong, A. V. Fedorov, H. Lin, A. Bansil, D. Grauer, Y. S. Hor, R. J. Cava, and M. Z. Hasan,
	Nature (London) \textbf{460}, 1101 (2009).
	
	\bibitem{ARPES_5} Y. L. Chen, J. G. Analytis, J.-H. Chu, Z. K. Liu, S.-K. Mo, X. L. Qi, H. J. Zhang, D. H. Lu, X. Dai Z. Fang, S. C. Zhang, I. R. Fisher, Z. Hussain, and Z.-X. Shen,
	Science \textbf{325}, 178 (2009).
	
	\bibitem{ARPES_1}  C. Chen, S. He, H. Weng, W. Zhang, L. Zhao, H. Liu, X. Jia, D. Mou, S. Liu, J. He, Y. Peng, Y. Feng, Z. Xie, G. Liu, X. Dong, J. Zhang, X. Wang, Q. Peng, Z. Wang, S. Zhang, F. Yang, C. Chen, Z. Xu, X. Dai, Z. Fang, and X. J. Zhou,
	PNAS \textbf{109}, 3694 (2012).
	
	\bibitem{ARPES_6} Z.-H. Pan, E. Vescovo, A. V. Fedorov, G. D. Gu, and T. Valla,
	Phys. Rev. B \textbf{88}, 041101 (2013). 
	
	\bibitem{timeARPES} M. Hajlaoui, E. Papalazarou, J. Mauchain, Z. Jiang, I. Miotkowski, Y.P. Chen, A. Taleb-Ibrahimi, L. Perfetti, and M. Marsi
	Eur. Phys. J. Special Topics \textbf{222}, 1271
	(2013).
	
	\bibitem{ARPES_4}  Y. Liu,	Y. Y. Li,	S. Rajput,	D. Gilks,	L. Lari,	P. L. Galindo,	M. Weinert,	V. K. Lazarov, and L. Li
	Nat. Phys. \textbf{10}, 294
	(2014).
	
	\bibitem{2ppe1} J.  Reimann, J.  Gudde, K.  Kuroda, E. V. Chulkov, and U. Hofer
	Phys. Rev. B. \textbf{90}, 081106 (2014).
	
	\bibitem{2ppe3} S. Zhu, Y. Ishida, K. Kuroda, K. Sumida, M. Ye, J. Wang, H. Pan, M. Taniguchi, S. Qiao, S. Shin, and A. Kimura
	Scient. Rep. \textbf{5}, 13213 (2015).	
	
	\bibitem{2ppe2} K. Kuroda, J. Reimann, J. G\"{u}dde, and U. H\"{o}fer
	Phys. Rev. Lett. \textbf{116}, 076801 (2016).
	
	\bibitem{STM1} G. Zhang, H. Qin, J. Teng, J. Guo, Q. Guo, X. Dai, Z. Fang, and K. Wu
	Appl. Phys. \textbf{95}, 053114 (2009).
	
	\bibitem{STM2} Z.Y. Wang, H.D. Li,  X. Guo , W.K. Ho, and M.H. Xie
	J. Cryst. Growth \textbf{334}, 96 
	(2011).
	
	\bibitem{STM3} Z. Alpichshev, R. R. Biswas, A. V. Balatsky, J. G. Analytis, J.-H. Chu, I. R. Fisher, and A. Kapitulnik
	Phys. Rev. Lett. \textbf{108}, 206402 (2012).
	
	\bibitem{Samarath}  A. Richardella, A. Kandala, J. Sue Lee, and N. Samarth
	Appl. Phys. Lett Mater \textbf{3}, 083303 (2015).
	
	\bibitem{SEM} M. Zhang, Z. Wei, R. Jin, Y. Ji, Y. Yan, X. Pu, X. Yang, and Y. Zhao, 
	Thin Solid Films \textbf{603}, 289 (2016). 
	
	\bibitem{AFM} G. Bendt, S. Zastrow, K. Nielsch, P. S. Mandal, J. Sanchez-Barriga, O. Rader, and S. Schulz, 
	J. Mater. Chem. A, {\bf 2}, 8215 (2014).
	
	\bibitem{XRD} H. Cao, R. Venkatasubramanian, C. Liu, J. Pierce, H. Yang, M. Z. Hasan, Y. Wu, and Y. P. Chen, 
	Appl. Phys. Lett. {\bf 101}, 162104 (2012).
	
	\bibitem{Hamh} S. Y. Hamh et al., 
	Appl. Phys. Lett., \textbf{108}  051609 (2016).
	
	\bibitem{Basov} B. C. Chapler, K. W. Post, A. R. Richardella, 
	J. S. Lee, J. Tao, N. Samarth, and D. N. Basov,
	Phys. Rev. B \textbf{89}, 235308 (2014).
	
	\bibitem{Loosdrecht}  N. Borgwardt, J. Lux, I. Vergara, Zhiwei Wang, A. A. Taskin, Kouji Segawa, P. H. M. van Loosdrecht, Y. Ando, A. Rosch, and M. Gr\"{u}ninger
	Phys. Rev. B, \textbf{93} 245149 (2016).
	
	
	\bibitem{Ganichev2003} S. D. Ganichev and W. Prettl, 
	topical review, J. Phys.: Condens. Matter, \textbf{15}, R935 (2003).
	
	\bibitem{book} S. D. Ganichev and  W. Prettl, 
	\textit{Intense Terahertz Excitation of Semiconductors}
	(Oxford Univ. Press, Oxford, 2006).
	
	\bibitem{Olbrich2014} P. Olbrich, L. E. Golub, T. Herrmann, S.N. Danilov, H. Plank, V.V. Bel'kov, G. Mussler, Ch. Weyrich, C. M. Schneider, J. Kampmeier, D. Gr\"{u}utzmacher, L. Plucinski, M. Eschbach, and S. D. Ganichev, 
	Phys. Rev. Lett. \textbf{113}, 096601 (2014).
	
	\bibitem{Kastl2015_1} Ch. Kastl, Ch. Karnetzky, H. Karl, and A. W. Holleitner, 
	Nature Comm. \textbf{6}, 6617 (2015).
	
	\bibitem{Braun2015} L. Braun, G. Mussler, A. Hruban, M. Konczykowski, M. Wolf, T. Schumann, M. M\"{u}nzenberg, L. Perfetti, T. Kampfrath, 
	arXiv:1511.00482 (2015).
	
	\bibitem{Plank2016} H. Plank, L. E. Golub, S. Bauer, V.V. Bel’kov, T. Herrmann, P. Olbrich, M. Eschbach, L. Plucinski, C. M. Schneider, J. Kampmeier, M. Lanius, G. Mussler, D. Gr\"{u}tzmacher, and S. D. Ganichev
	Phys. Rev. B \textbf{93}, 125434  (2016).
	
	\bibitem{Borisova2012} S. Borisova, J. Krumrain, M. Luysberg, G. Mussler, and D. Gr\"{u}tzmacher,
	Cryst. Growth Des. \textbf{12}, 6098
	(2012).
	
	\bibitem{Plucinski2013} L. Plucinski, A. Herdt, S. Fahrendorf, G. Bihlmayer, G. Mussler, S. D\"{o}ring, J. Kampmeier, F. Matthes, D. E. Bürgler, D. Gr\"{u}tzmacher, S. Bl\"{u}gel, and C. M. Schneider, 
	Appl. Phys. \textbf{113}, 053706 (2013).
	
	\bibitem{Borisova2013} S. Borisova, J. Kampmeier, M. Luysberg, G. Mussler and D. Gr\"{u}tzmacher,
	Appl. Phys. Lett. \textbf{103}, 081902 (2013).
	
	\bibitem{Kampmeier2015} J. Kampmeier, S. Borisova, L. Plucinski, M. Luysberg, G. Mussler, and D. Gr\"{u}tzmacher, 
	Cryst. Growth Des., \textbf{15}, 390 (2015).
	
	\bibitem{ternaries} C. Weyrich, M. Dr\"ogeler, J. Kampmeier, M. Eschbach, G. Mussler, T. Merzenich, T. Stoica, I. E. Batov, J. Schubert, L. Plucinski, B. Beschoten, C. M. Schneider, C. Stampfer, D. Gr\"utzmacher, and Th. Sch\"apers, 
	arxiv 1511.00965v2 (2015).
	
	
	\bibitem{removalSiGe}  
	S.D. Ganichev et al., 
	Phys. Rev. B \textbf{66}, 075328   (2002). 
	
	\bibitem{Resonantinversion2003} S.D. Ganichev, V. V. Bel'kov, Petra Schneider, E. L. Ivchenko, S. A. Tarasenko,  W. Wegscheider, D. Weiss,    D. Schuh, E. V. Beregulin  and  W. Prettl, 
	Phys. Rev. B {\bf 68}, 035319 (2003).
	
	\bibitem{SGEopt2003} S. D. Ganichev, Petra Schneider, V. V. Bel'kov, E. L. Ivchenko, S. A. Tarasenko, W. Wegscheider, D. Weiss, D. Schuh, B. N. Murdin, P. J. Phillips,  C. R. Pidgeon, D. G. Clarke, M. Merrick,  P. Murzyn,  E. V. Beregulin, and W. Prettl,
	Phys. Rev. B. R {\bf 68}, 081302 (2003).
	
	\bibitem{Chongyun} 
	Chongyun Jiang et a., 
	Phys. Rev. B \textbf{84}, 125429   (2011).
	
	\bibitem{Lechner2009}
	V. Lechner, L. E. Golub, P. Olbrich, S. Stachel, D. Schuh, W. Wegscheider, V. V. Bel'kov, and S. D. Ganichev,
	Appl. Phys. Lett. \textbf{94}, 242109 (2009).
	
	\bibitem{Drexler2013} C. Drexler,  S. A. Tarasenko, P. Olbrich, J. Karch, M. Hirmer, F. M\"{u}ller, M. Gmitra, J. Fabian, R. Yakimova, S. Lara-Avila, S. Kubatkin, and S. D. Ganichev,
	Nature Nanotechnology \textbf{8}, 104 (2013).
	
	\bibitem{ratchet2009} P. Olbrich, E. L. Ivchenko, T. Feil, R. Ravash, S. D. Danilov, J. Allerdings, D. Weiss, and S. D. Ganichev,
	Phys. Rev. Lett. \textbf{103}, 090603 (2009).
	
	\bibitem{helix2012} M. Kohda, V. Lechner, Y. Kunihashi, T. Dollinger, P. Olbrich, C. Sch\"{o}nhuber,  I. Caspers, V. V. Bel'kov, L. E. Golub, D. Weiss, K. Richter, J. Nittaand S. D. Ganichev,
	Phys. Rev. B Rapid Communic. \textbf{86}, 081306 (2012).
	
	\bibitem{Ganichev84p20} S. D. Ganichev, Y. V. Terent'ev, and I. D. Yaroshetskii, 
	Pisma Zh. Tekh. Fiz. \textbf{11}, 46 (1985) [Sov. Tech. Phys. Lett. \textbf{11}, 20 (1989)].
	
	\bibitem{BelkovSSTlateral} V. V. Bel'kov,  and S. D. Ganichev,
	Semicond. Sci. Technol. \textbf{23}, 114003 (2008).
	
	\bibitem{ratchet2011} P. Olbrich, J. Karch, E. L. Ivchenko, J. Kamann, B. M{\"a}rz, M. Fehrenbacher, D. Weiss, and S. D. Ganichev,
	Phys. Rev. B \textbf{83}, 165320 (2011).
	
	\bibitem{Ganichev1999} S. D. Ganichev
	Physica B \textbf{273-274}, 737 (1999).
	
	\bibitem{Ziemann2000}  E. Ziemann, S. D. Ganichev, I. N. Yassievich, V. I. Perel, and W. Prettl,
	J. Appl. Phys. {\bf 87}, 3843 (2000).
	
	\bibitem{NatPhys09} H. Zhang, C.-X. Liu, X.-L. Qi, X. Dai, Z. Fang, and S.-C. Zhang,
	Nature Phys. \textbf{5}, 438 (2009).
	
	
	\bibitem{footnoteT1} Note that for quasi-elastic scattering by phonons we obtain the expression 
	which differs from the equation above by a constant factor of order of unity only.
	
	\bibitem{footnoteT3}
	Due to the fact that in the applied frequency range the 
	condition $\omega _{\rm tr}\tau \gg 1$ is satisfied we are limited 
	here to the statement that the room temperature mobility of the surface states 
	is at least higher that 1000\,cm$^2$/Vs. However, photogalvanic experiments in a 
	spectral range substantially extended to lower frequencies, e.g. applying Gunn diode or backwardwave
	oscillator, would allows one to measure the electron mobility.
	In this regime the LPGE photocurrent magnitude $A \propto \sigma(\omega)$
	will change its behaviour from $A~\propto (\omega\tau_{tr})^{-2}$ to the 
	frequency independent one:  a function allowing to obtain $\tau_{tr}$
	with rather high accuracy.
	
	\bibitem{footnoteT2} In the case contacts or edges are illuminated the signal becomes polarization independent and increases by three orders of magnitude compared to photogalvanic effect in middle part of the sample.  The assignment of this signal which can be caused 
	by e.g. $p-n$ junctions or thermal effects in inhomogeneous area close to contacts 
	is out of scope of this paper.
	
\end{thebibliography}
\end{document}